\documentclass{PoS-hep}
\usepackage{amsmath}
\usepackage{bm}
\usepackage{epsfig}
\usepackage{graphics}
\usepackage{upgreek}
\usepackage{amsfonts}
\usepackage{amsbsy}
\usepackage{amscd}
\usepackage{bbm}
\usepackage{eufrak}
\usepackage{cite}

\renewenvironment{subequations}{%
\refstepcounter{equation}%
\setcounter{parentequation}{\value{equation}}%
  \setcounter{equation}{0}
  \ignorespaces
}{%
  \setcounter{equation}{\value{parentequation}}%
  \ignorespacesafterend
}
\newcommand{\eeq}{\end{equation}}
\newcommand{\beq}{\begin{equation}}
\newcommand{\ba}{\begin{array}}
\newcommand{\ea}{\end{array}}
\newcommand{\bea}{\begin{eqnarray}}
\newcommand{\eea}{\end{eqnarray}}

\newcommand{\beqs}{\begin{subequations}}
\newcommand{\eeqs}{\end{subequations}}
\newcommand{\eec}{\end{center}}
\newcommand{\bec}{\begin{center}}
\newcommand{\eem}{\end{matrix}}
\newcommand{\bem}{\begin{matrix}}
\newcommand{\Eref}[1]{Eq.~(\ref{#1})}
\newcommand{\Sref}[1]{Sec.~\ref{#1}}
\newcommand{\Fref}[1]{Fig.~\ref{#1}}
\newcommand{\Tref}[1]{Table~\ref{#1}}
\newcommand{\cref}[1]{Ref.~\cite{#1}}

\newcommand\eqs[2]{Eqs.~(\ref{#1}) and (\ref{#2})}

\newcommand\eqss[3]{Eqs.~(\ref{#1}), (\ref{#2}) and (\ref{#3})}

\newcommand{\sEref}[2]{Eq.~(\ref{#1}{\ftn\sf {#2}})}
\newcommand{\sFref}[2]{Fig.~\ref{#1}-{\ftn\sf ({#2})}}

\newcommand{\ftn}{\footnotesize}

\newcommand{\GeV}{{\mbox{\rm GeV}}}

\newcommand{\etal}{{\it et al.\/}}

\def\to{\rightarrow}

\def\lf{\left(}
\def\rg{\right)}
\newcommand\vev[1]{\langle {#1} \rangle}

\newcommand{\Vhi}{\ensuremath{\widehat V_{\rm IG}}}
\newcommand{\Hhi}{\ensuremath{\widehat H_{\rm IG}}}
\newcommand{\Ohi}{\ensuremath{\Omega}}
\newcommand{\Omg}{\ensuremath{\Omega}}
\newcommand{\Khi}{\ensuremath{K}}
\newcommand{\Whi}{\ensuremath{W}}
\newcommand{\Vhio}{\ensuremath{\widehat V_{\rm IG0}}}

\newcommand{\mP}{\ensuremath{m_{\rm P}}}

\newcommand{\Qef}{\ensuremath{\Lambda_{\rm UV}}}

\def\openone{\leavevmode\hbox{\small1\kern-3.8pt\normalsize1}}

\newcommand{\ft}{\ensuremath{f_{W}}}
\newcommand{\kx}{\ensuremath{k_S}}

\newcommand{\Fcr}{\ensuremath{\Omega_{\rm H}}}
\newcommand{\fr}{\ensuremath{f_{R}}}
\newcommand{\fk}{\ensuremath{\Omega_{\rm H}}}

\newcommand{\re}{\ensuremath{e_n}}

\newcommand{\fpp}{\ensuremath{f_{W}}}
\newcommand{\fsp}{\ensuremath{f_{S\Phi}}}
\newcommand{\Fk}{\ensuremath{\Omega_{\rm K}}}
\newcommand{\ks}{\ensuremath{k_S}}
\newcommand{\ksp}{\ensuremath{k_{S\Phi}}}
\newcommand{\kpp}{\ensuremath{k_{\Phi}}}
\newcommand{\kns}{\ensuremath{k_{\sf NS}}}

\newcommand{\ck}{\ensuremath{c_{R}}}

\newcommand{\ns}{\ensuremath{n_{\rm s}}}
\newcommand{\as}{\ensuremath{a_{\rm s}}}
\newcommand{\As}{\ensuremath{A_{\rm s}}}
\newcommand{\rcc}{\ensuremath{R}}
\newcommand{\rce}{\ensuremath{\widehat{R}}}
\newcommand{\Ve}{\ensuremath{\widehat{V}}}
\newcommand{\He}{\ensuremath{{\what H}}}
\newcommand{\Ne}{\ensuremath{{\what N}}}
\newcommand{\Ns}{\ensuremath{\what N_{\star}}}

\newcommand{\dphi}{\ensuremath{\what{\delta\phi}}}
\newcommand{\dph}{\ensuremath{\delta\phi}}

\newcommand{\what}{\ensuremath{\widehat}}

\def\aal{{\bar\alpha}}
\def\bbet{{\bar\beta}}
\def\al{{\alpha}}
\def\bt{{\beta}}

\def\th{{\theta}}

\newcommand{\Trh}{\ensuremath{T_{\rm rh}}}
\newcommand{\sg}{\ensuremath{\phi}}

\newcommand{\sgx}{\ensuremath{\phi_\star}}
\newcommand{\sgf}{\ensuremath{\phi_{\rm f}}}
\newcommand{\xsg}{\ensuremath{\phi}}
\newcommand{\xst}{\ensuremath{\phi_{\star}}}

\newcommand{\ld}{\ensuremath{\lambda}}

\newcommand{\se}{\ensuremath{\widehat \phi}}
\newcommand{\sex}{\ensuremath{\widehat{\phi}_\star}}

\newcommand{\geu}{\ensuremath{\widehat g}}
\newcommand{\eph}{\ensuremath{\widehat \epsilon}}

\def\trns{transplanckian}
\def\Ka{K\"{a}hler potential}
\def\Km{K\"{a}hler manifold}

\def\sub{subplanckian}
\def\FHI{IGI~}

\def\bcp{B{\sc\small icep2}/{\it Keck Array}}

\newcommand{\plk}{\emph{Planck}}

\title{Induced-Gravity Inflation in Supergravity
Confronted with \emph{Planck} 2015 \& {\bfseries\scshape
Bicep2}/{\slshape Keck Array}}

\ShortTitle{IGI in SUGRA Confronted with {\it Planck} 2015 \& {\sc
B\small icep2}/{\it Keck Array}}

\author{\speaker{C. Pallis}\\
Departament de F\'isica Te\`orica and IFIC,\\
Universitat de Val\`encia-CSIC, \\ E-46100 Burjassot, SPAIN\\
E-mail: \email{cpallis@ific.uv.es}}

\abstract{Supersymmetric versions of induced-gravity inflation are
formulated within Supergravity (SUGRA) employing two gauge singlet
chiral superfields. The proposed superpotential is uniquely
determined by applying a continuous $R$ and a discrete
$\mathbb{Z}_2$ symmetry. We also employ a logarithmic \Ka\
respecting the symmetries above and including all the allowed
terms up to fourth order in powers of the various fields. When the
\Km\ exhibits a no-scale-type symmetry, the model predicts
spectral index $\ns\simeq0.963$ and tensor-to-scalar
$r\simeq0.004$. Beyond no-scale SUGRA, $\ns$ and $r$ depend
crucially on the coefficient $\ksp$ involved in the fourth order
term, which mixes the inflaton $\Phi$ with the accompanying
non-inflaton superfield $S$ in the \Ka, and the prefactor
encountered in it. Increasing slightly the latter above $(-3)$, an
efficient enhancement of the resulting $r$ can be achieved putting
it in the observable range favored by the \plk\ and \bcp\ results.
In all cases, imposing a lower bound on the parameter $\ck$,
involved in the coupling between the inflaton and the Ricci scalar
curvature, inflation can be attained for \sub\ values of the
inflaton while the corresponding effective theory respects the
perturbative unitarity.
\\ \\
{\sl\bfseries Published in}~~{PoS  CORFU {\bf 2014}, 156 (2015)}.
}

\FullConference{Proceedings of the Corfu Summer Institute 2014 \\
                 3-21 September 2014\\
                 Corfu, Greece}

\begin{document}

\section{Introduction}

\emph{Induced-gravity inflation} ({\sf\ftn IGI}) \cite{old} is a
subclass of non-minimal inflationary models in which inflation is
driven in the presence of a non-minimal coupling function between
the inflaton field and the Ricci scalar curvature and the Planck
mass is determined by the \emph{vacuum expectation value}
({\sf\ftn v.e.v}) of the inflaton at the end of the slow roll. As
a consequence, IGI not only is attained even for subplanckian
values of the inflaton -- thanks to the strong enough
aforementioned coupling -- but also the corresponding effective
theory remains valid up to the Planck scale \cite{R2r, gian}. In
this talk we focus on the implementation of IGI within
\emph{Supergravity} ({\sf\ftn SUGRA}) \cite{nIG, rena} revising
and updating the findings of \cref{nIG} in the light of the recent
joint analysis \cite{plin,gws} of \plk\ and \bcp\ results.

Below, in Sec.~\ref{fhim}, we describe the generic formulation of
IGI in SUGRA. The established in Sec.~\ref{fhi} inflationary
models are investigated in Sec.~\ref{res}. The \emph{ultraviolet}
({\sf\ftn UV}) behavior of these models is analyzed in
Sec.~\ref{fhi3}. Our conclusions are summarized in Sec.~\ref{con}.
Throughout the text, the subscript $,\chi$ denotes derivation
\emph{with respect to} ({\ftn\sf w.r.t}) the field $\chi$; charge
conjugation is denoted by a star, and we use units where the
reduced Planck scale $\mP = 2.435\cdot 10^{18}~\GeV$ is set equal
to unity.

\section{Embedding IGI in SUGRA}\label{fhim}

According to the scheme proposed in \cref{nIG}, the implementation
of IGI in SUGRA requires at least two singlet superfields, i.e.,
$z^\al=\Phi, S$, with $\Phi$ ($\al=1$) and $S$ ($\al=2)$ being the
inflaton and a stabilized field respectively. The superpotential
$W$ of the model has the form
\beq\label{Whi} W= {\ld\over\ck} S\lf
\fk-1/2\rg~~~\mbox{with}~~~\fk(\Phi)=\ck
{\Phi^2}+\sum_{k=1}^\infty\lambda_{k}{\Phi^{4k}}\,,\eeq
which is {\sf\ftn (i)} invariant under the action of a global
$\mathbb{Z}_2$ discrete symmetry, i.e., \beq W \to\ \,
W~~~\mbox{for}~~~\Phi\ \to\ -\Phi~~~\mbox{and}~~~ S\to S\label{Zn}
\eeq and {\sf\ftn (ii)} consistent with a continuous $R$ symmetry
under which \beq W \to\ e^{i\varphi}\, W~~~\mbox{for}~~~S\ \to\
e^{i\varphi}\,S~~~\mbox{and}~~~\fk\ \to\ \fk\,.\label{Rsym} \eeq
Confining ourselves to $\Phi<1$ and assuming relatively low
$\lambda_{k}$'s we hereafter neglect the second term in the
definition of $\fk$ in \Eref{Whi}. The \emph{Supersummetric}
({\sf\ftn SUSY}) F-term scalar potential obtained from $W$ in
\Eref{Whi} is
\beq V_{\rm F}= \ld^2\left| \fk- 1/2\right|^2/\ck^2 +
\ld^2|S\Omega_{{\rm H},\Phi}|^2/\ck^2, \label{VF}\eeq
where the complex scalar components of $\Phi$ and $S$ are denoted
by the same symbol. From \Eref{VF}, we find that the SUSY vacuum
lies at the direction
\beq \vev{S}=0~~~\mbox{and}~~~ \vev{\fk}=1/2,\label{vevs} \eeq
where we take into account that the phase of $\Phi$, $\arg\Phi$,
is stabilized to zero during and after IGI. If $\fk$ is the
holomorphic part of the frame function $\Omega$ and dominates it,
\Eref{vevs} assures a transition to the conventional Einstein
gravity realizing, thereby, the idea of induced gravity
\cite{old}.

To combine this idea with an inflationary setting we have to
define a suitable relation between $\Omg$ and the \Ka\ $K$ so as
the scalar potential far away from the SUSY vacuum to admit
inflationary solutions. To this end, we focus on \emph{Einstein
frame} ({\sf\ftn EF}) action for $z^\al$'s within SUGRA
\cite{linde1} which is written as
\beq\label{Saction1} {\sf S}=\int d^4x \sqrt{-\what{
\mathfrak{g}}}\lf-\frac{1}{2} \rce +K_{\al\bbet}\geu^{\mu\nu}
\partial_\mu z^\al \partial_\nu z^{*\bbet}-\Ve\rg, \eeq
where $\Ve$ is the F--term SUGRA scalar potential given below,
summation is taken over the scalar fields $z^\al$,
$K_{\al\bbet}={K_{,z^\al z^{*\bbet}}}$ with
$K^{\bbet\al}K_{\al\bar \gamma}=\delta^\bbet_{\bar \gamma}$,
$\widehat{\mathfrak{g}}$ is the determinant of the EF metric
$\geu_{\mu\nu}$. If we perform a conformal transformation defining
the \emph{Jordan frame} ({\sf\ftn JF}) metric $g_{\mu\nu}$ through
the relation
\beq \label{weyl}
\geu_{\mu\nu}=-\frac{\Omega}{3(1+n)}g_{\mu\nu}~~\Rightarrow~~\left\{\bem
\sqrt{-\what{ \mathfrak{g}}}={\Omega^2\over
9(1+n)^2}\sqrt{-\mathfrak{g}}~~~\mbox{and}~~~
\geu^{\mu\nu}=-{3(1+n)\over\Omega}g^{\mu\nu}, \hfill \cr
\rce=-{3(1+n)\over\Omega}\left(\rcc-\Box\ln \Omega+3g^{\mu\nu}
\partial_\mu \Omega\partial_\nu \Omega/2\Omega^2\right) \hfill
\cr\eem
\right.\eeq
where $n$ is a dimensionless (small in our approach) parameter
which quantifies the deviation from the standard set-up
\cite{linde1}, ${\sf S}$ is written in the JF as follows
\beq {\sf S}=\int d^4x \sqrt{-\mathfrak{g}}\lf\frac{\Omega
\rcc}{6(1+n)}+\frac{\Omega\partial_\mu\Omega\partial^\mu\Omega}{4(1+n)}
-\frac{1}{(1+n)}\Omega K_{\al{\bbet}}\partial_\mu z^\al
\partial^\mu z^{*\bbet}-V  \rg\label{action2}\eeq
with $V ={\Omega^2\Ve/9(1+n)^2}$ being the JF potential in
\Eref{VF}. If we specify the following relation between $\Omega$
and $K$,
\beq-\Omega/3(1+n)
=e^{-K/3(1+n)}\>\Rightarrow\>K=-3(1+n)\ln\lf-\Omega/3(1+n)\rg,\label{Omg1}\eeq
and employ the definition  \cite{linde1} of the purely bosonic
part of the on-shell value of the auxiliary field
\beq {\cal A}_\mu =i\lf K_\al\partial_\mu z^\al-K_\aal\partial_\mu
z^{*\aal}\rg/6, \label{Acal1}\eeq
we arrive at the following action
\beq {\sf S}=\int d^4x
\sqrt{-\mathfrak{g}}\lf\frac{\Omega\rcc}{6(1+n)}+
\lf\Omega_{\al{\bbet}}-\frac{n\Omega_{\al}\Omega_{\bbet}}{(1+n)\Omega}\rg\partial_\mu
z^\al \partial^\mu z^{*\bbet}- \frac{\Omega{\cal A}_\mu{\cal
A}^\mu}{(1+n)^3}-V \rg, \label{Sfinal}\eeq
where ${\cal A}_\mu$ in \Eref{Acal1} takes the form
\beq {\cal A}_\mu =-i(1+n)\lf \Omega_\al\partial_\mu
z^\al-\Omega_\aal\partial_\mu z^{*\aal}\rg/2\Omega\,.
\label{Acal}\eeq
It is clear from \Eref{Sfinal} that ${\sf S}$ exhibits non-minimal
couplings of the $z^\al$'s to $\rcc$. However, $\Omega$ also
enters the kinetic terms of the $z^\al$'s. To separate the two
contributions we split $\Omega$ into two parts
\beqs\beq -\Omega/3(1+n)=\fk(\Phi)+{\fk}^*(\Phi^*)-\Fk\lf|\Phi|^2,
|S|^2\rg/3(1+n), \label{Omg}\eeq
where $\Fk$ is a dimensionless real function including the kinetic
terms for the $z^\al$'s and takes the form
\beq \label{Fkdef} \Fk\lf|\Phi|^2, |S|^2\rg= {\kns
|\Phi|^2+|S|^2}\,-\, 2\lf\kx|S|^4+\kpp|\Phi|^4+\ksp
|S|^2|\Phi|^2\rg\,\eeq\eeqs
with coefficients $\kns, \kx, \kpp$ and $\ksp$ of order unity. The
fourth order term for $S$ is included to cure the problem of a
tachyonic instability occurring along this direction
\cite{linde1}, and the remaining terms of the same order are
considered for consistency -- the factors of $2$ are added just
for convenience. On the other hand, $\Fcr$ in \Eref{Omg} is a
dimensionless holomorphic function which, for $\Fcr>\Fk$,
represents the non-minimal coupling to gravity -- note that
$\Omega_{\al{\bbet}}$ is independent of $\Fcr$ since $\Omg_{{\rm
H},z^\al z^{*\bbet}}=0$. If $\arg\Phi$ is stabilized to zero, then
$\fk=\fk^*$ and from \eqs{Sfinal}{Omg} we deduce that \Eref{vevs}
recovers the conventional term of the Einstein gravity at the SUSY
vacuum implementing thereby the idea of induced gravity. The
choice $n\neq0$, although not standard, is perfectly consistent
with the set-up of non-minimal inflation \cite{linde1} since the
only difference occurring for $n\neq0$ is that the $z^\al$'s do
not have canonical kinetic terms in the JF due to the term
proportional to $\Omg_\al\Omg_\bbet\neq\delta_{\al\bbet}$ in
\Eref{Sfinal}. This fact does not cause any problem since the
canonical normalization of $\Phi$ keeps its strong dependence on
$\ck$, whereas $S$ becomes heavy enough during IGI and so it does
not affect the dynamics -- see \Sref{fgi1}.

In conclusion, through \Eref{Omg1} the resulting  \Ka\ is
\beq  \Khi=-3(1+n)\ln\lf{\ck}\lf\Phi^2+\Phi^{*2}\rg-{
|S|^2+\kns|\Phi|^2\over3(1+n)}+2{\kx|S|^4+\kpp|\Phi|^4+\ksp
|S|^2|\Phi|^2\over3(1+n)}\rg.\label{Kolg}\eeq
We set $\kns=1$ throughout, except for the case of no-scale SUGRA
which is defined as follows:
\beq n=0,~~~\kns=0~~~\mbox{and}~~~\ksp=\kpp=0\,.\label{nsks}\eeq
This arrangement, inspired by the early models of soft SUSY
breaking \cite{noscale,R2r}, corresponds to the \Km\ $SU(2,1)/
SU(2)\times U_R(1)\times \mathbb{Z}_2$ with constant curvature
equal to $-2/3$. In practice, these choices highly simplify the
realization of IGI, rendering it more predictive thanks to a lower
number of the remaining free parameters.

\section{Inflationary Set-up}\label{fhi}

In this section we describe -- in Sec.~\ref{fgi1} -- the
derivation of the inflationary potential of our model and then --
in \Sref{fhi2} -- we exhibit a number of observational and
theoretical constraints imposed.

\subsection{Inflationary Potential}\label{fgi1}

The EF F--term (tree level) SUGRA scalar potential $\Ve$,
encountered in \Eref{Saction1}, is obtained from $\Whi$ and $\Khi$
in \eqs{Whi}{Kolg} respectively by applying (for $z^\al=\Phi,S$)
the well-known formula
\beq \Ve=e^{\Khi}\left(K^{\al\bbet}D_\al W D^*_\bbet W^* -3{\vert
W\vert^2}\right)~~~\mbox{with}~~~D_\al W=W_{,z^\al}
+K_{,z^\al}W.\label{Vsugra}\eeq
Along the inflationary track determined by the constraints
\beq \label{inftr} S=\Phi-\Phi^*=0,~\mbox{or}~~s=\bar s=\th=0\eeq
if we express $\Phi$ and $S$ according to the standard
parametrization
\beq \Phi=\:{\phi\,e^{i \th}}/{\sqrt{2}}~~~\mbox{and}~~~S=\:(s
+i\bar s)/\sqrt{2}\,,\label{cannor} \eeq
the only surviving term in \Eref{Vsugra} is
\beq\Vhio=\Ve(\th=s=\bar s=0)=e^{K}K^{SS^*}\,
|W_{,S}|^2=\frac{\ld^2|2\fk-1|^2}{4\ck^2\fsp\fr^{2+3n}}\,\cdot\label{Vhig}\eeq
Here we take into account that
\beqs\beq
e^{K}=\fr^{-3(1+n)}~~~\mbox{and}~~~K^{SS^*}={\fr/\fsp},\label{Vhigg}\eeq
where the functions $\fr$ and $\fsp$ are defined along the
direction in \Eref{inftr} as follows:
\bea \label{frsp}
\fr=-\frac{\Ohi}{3(1+n)}=\ck{\xsg^2}-{\kns\xsg^2-\kpp\xsg^4\over6(1+n)}~~~\mbox{and}~~~
\fsp=\Omega_{,SS^*}=1-\ksp\xsg\,.\eea\eeqs
Given that $\fsp\ll\fr\simeq2\fk$ with $\ck\gg1$, $\Vhio$ in
\Eref{Vhig} is roughly proportional to $\xsg^{-6n}$. Therefore, an
inflationary plateau emerges for $n=0$ and a chaotic-type
potential (bounded from below) is generated for $n<0$. More
specifically, $\Vhio$ and the corresponding EF Hubble parameter,
$\He_{\rm IG}$, can be cast in the following form:
\beq \Vhio=\frac{\ld^2\ft^2\xsg^{-6n}}{4\ck^2\xsg^4\fsp}
\lf\ck-\frac{f_{\phi\phi}}{6(1+n)}\rg^{-(2+3n)} \simeq\frac{\ld^2
\mP^4\xsg^{-6n}}{4\fsp\ck^{2+3n}}\,~~~\mbox{and}~~~\He_{\rm
IG}={\Vhio^{1/2}\over\sqrt{3}}\simeq{\ld\xsg^{-3n}\over2\sqrt{3\fsp}\ck^{1+3n/2}}\,,\label{3Vhiom}\eeq
where we introduce the functions $f_{\phi\phi}=1-\kpp\xsg^2$ and
$\ft=1-\ck\xsg^2$.

The stability of the configuration in \Eref{inftr} can be checked
verifying the validity of the conditions
\beq {\partial \Ve/\partial\what\chi^\al}=0~~~ \mbox{and}~~~\what
m^2_{ \chi^\al}>0~~~\mbox{with}~~~\chi^\al=\th,s,\bar
s,\label{Vcon} \eeq
where $\what m^2_{\chi^\al}$ are the eigenvalues of the mass
matrix with elements $\what
M^2_{\al\bt}={\partial^2\Ve/\partial\what\chi^\al\partial\what\chi^\beta}$
and hat denotes the EF canonically normalized fields defined by
the kinetic terms in \Eref{Saction1} as follows
\beqs\beq \label{K3} K_{\al\bbet}\dot z^\al \dot
z^{*\bbet}=\frac12\lf\dot{\se}^{2}+\dot{\what
\th}^{2}\rg+\frac12\lf\dot{\what s}^2 +\dot{\what{\overline
s}}^2\rg,\eeq
where the dot denotes derivation w.r.t the JF cosmic time and the
hatted fields read
\beq  \label{Jg} {d\widehat \sg/
d\sg}=\sqrt{K_{\Phi\Phi^*}}=J\simeq{\sqrt{6(1+n)}/\xsg},~~~
\what{\th}= J\,\th\xsg~~~\mbox{and}~~~(\what s,\what{\bar
s})=\sqrt{K_{SS^*}} {(s,\bar s)}\,,\eeq\eeqs
where $K_{SS^*}\simeq1/\ck\xsg^2$ -- cf. \eqs{Vhigg}{frsp}. The
spinors $\psi_\Phi$ and $\psi_S$ associated with $S$ and $\Phi$
are normalized similarly, i.e.,
$\what\psi_{S}=\sqrt{K_{SS^*}}\psi_{S}$ and
$\what\psi_{\Phi}=\sqrt{K_{\Phi\Phi^*}}\psi_{\Phi}$. Integrating
the first equation in \Eref{Jg} we can identify the EF field as
\beq \se=\se_{\rm
c}+\sqrt{6(1+n)}\ln\lf{\sg}/{\vev{\sg}}\rg~~~\mbox{with}
~~~\vev{\phi}=1/{\sqrt{\ck}},\label{se1}\eeq
where $\se_{\rm c}$ is a constant of integration and we make use
of \eqs{Whi}{vevs}.

Upon diagonalization of $\what M^2_{\al\bt}$, we construct the
mass spectrum of the theory along the path of \Eref{inftr}. Taking
advantage of the fact that $\ck\gg1$ and the limits $\kpp\to0$ and
$\ksp\to0$ we find the expressions of the relevant masses squared,
arranged in \Tref{tab4}, which approach rather well the quite
lengthy, exact expressions taken into account in our numerical
computation. We have numerically verified that the various masses
remain greater than $\Hhi$ during the last $50$ e-foldings of
inflation, and so any inflationary perturbations of the fields
other than the inflaton are safely eliminated. They enter a phase
of oscillations about zero with reducing amplitude and so the
$\xsg$ dependence in their normalization -- see \Eref{Jg} -- does
not affect their dynamics. As usually -- cf. \cref{nMCI, R2r} --,
the lighter eignestate of $\what M^2_{\al\bt}$ is $\what
m^2_{{s}}$ which here can become positive and heavy enough for
$\kx\gtrsim0.05$ -- see \Sref{res2}.

\begin{table}[!t]
\renewcommand{\arraystretch}{1.4}
\bec\begin{tabular}{|c|c|l|}\hline
{\sc Fields} &{\sc Eingestates} & \hspace*{3.cm}{\sc Masses Squared}\\
\hline \hline
$1$ real scalar &$\what{\th}$ & $\what m^2_{\th}\simeq\ld^2
\lf 2-2\ck\xsg^2f_W+3n f_W^2\rg$\\ && $ /6(1 + n) \ck^{4 +3 n}\xsg^{2(2 + 3 n)} \simeq4\He_{\rm IG}^2$\\
$2$ real scalars &$\what{s},~\what{\bar s}$ & $\what m^2_{
s}=\ld^2 \lf2 -6n - \ck \xsg^2+ 12\ks (1+n)f_W^2\rg$
\\&&$/6 (1+n)\ck^{3(1+n)}\xsg^{2(1+3n)}$\\ \hline
$2$ Weyl spinors & $\what{\psi}_\pm={\what{\psi}_{\Phi}\pm
\what{\psi}_{S}\over\sqrt{2}}$& $\what m^2_{ \psi\pm}\simeq\ld^2
(2+3 n\fpp)^2/12(1 + n) \ck^{4 +3 n}\xsg^{2(2 + 3 n)}$
\\ \hline
\end{tabular}\eec
\renewcommand{\arraystretch}{1.}
\hfill \caption{\sl\small Mass spectrum along the inflationary
trajectory in Eg. (4.2).}\label{tab4}
\end{table}

Inserting, finally, the mass spectrum of the model in the
well-known Coleman-Weinberg formula, we calculate the one-loop
corrected inflationary potential
\beq\Vhi=\Vhio+{1\over64\pi^2}\lf \widehat m_{ \th}^4\ln{\widehat
m_{\th}^2\over\Lambda^2} +2 \widehat m_{ s}^4\ln{\widehat
m_{s}^2\over\Lambda^2}-4\widehat
m_{\psi_{\pm}}^4\ln{m_{\widehat\psi_{\pm}}^2\over\Lambda^2}\rg
,\label{Vhic}\eeq
where $\Lambda$ is a renormalization-group mass scale. We
determine it by requiring \cite{nMCI} $\Delta V(\sgx)=0$ with
$\Delta V=\Vhi-\Vhio$ the \emph{radiative corrections} ({\sf\ftn
RCs}) to $\Vhio$. To reduce the possible dependence of our results
on the choice of $\Lambda$, we confine ourselves to $\ld$'s and
$\kx$'s which do not enhance the RCs. Under these circumstances,
our results can be exclusively reproduced by using $\Vhio$.

\subsection{Inflationary Requirements}\label{fhi2}

Based on $\Ve_{\rm IG}$ in \Eref{Vhic} we can proceed to the
analysis of \FHI in the EF \cite{old}, employing the standard
slow-roll approximation. We have just to convert the derivations
and integrations w.r.t $\se$ to the corresponding ones w.r.t $\sg$
keeping in mind the dependence of $\se$ on $\sg$, \Eref{Jg}. In
our analysis we take into account the following observational and
theoretical requirements:

\paragraph{3.2.1} The number of e-foldings, $\Ns$, that
the scale $k_\star=0.05/{\rm Mpc}$ suffers during IGI has to be
adequate to resolve the horizon and flatness problems of standard
big bang, i.e., \cite{plin, R2r}
\begin{equation}
\label{Nhi}  \Ns=\int_{\se_{\rm f}}^{\se_{\star}}\, {d\se}\:
\frac{\Ve_{\rm IG}}{\Ve_{\rm IG,\se}} \simeq61.7+\ln{\what V_{\rm
IG}(\sgx)^{1/2}\over\what V_{\rm IG}(\sgf)^{1/3}}+ {1\over3}\ln
T_{\rm rh}+{1\over2}\ln{\fr(\sgx)\over\fr(\sgf)^{1/3}}\,,
\end{equation}
where $\sgx~[\se_\star]$ is the value of $\sg~[\se]$ when
$k_\star$ crosses outside the inflationary horizon and $\sg_{\rm
f}~[\se_{\rm f}]$ is the value of $\sg~[\se]$ at the end of IGI,
which can be found from the condition
\beq \label{sr} {\sf max}\{\widehat\epsilon(\sg_{\rm
f}),|\widehat\eta(\sg_{\rm
f})|\}=1,~~~\mbox{where}~~~\widehat\epsilon=
{1\over2}\left(\frac{\Ve_{\rm IG,\se}}{\Ve_{\rm
IG}}\right)^2~~~\mbox{and}~~~\widehat\eta=\frac{\Ve_{\rm
IG,\se\se}}{\Ve_{\rm IG}} \eeq
are the well-known slow-roll parameters and $\Trh$ is the reheat
temperature after IGI, which is taken $\Trh=4.1\cdot10^{-10}$
throughout. We also assume canonical reheating \cite{liddle} with
an effective equation-of-state parameter $w_{\rm re}=0$ and the
effective number of relativistic degrees of freedom at temperature
$\Trh$ is taken $g_{\rm rh}=228.75$ corresponding to the MSSM
spectrum.

\paragraph{3.2.2} The amplitude $\As$ of the power spectrum of the curvature perturbation
generated by $\sg$ at  $k_{\star}$ has to be consistent with
data~\cite{plin}
\begin{equation}  \label{Prob}
\sqrt{A_{\rm s}}=\: \frac{1}{2\sqrt{3}\, \pi} \; \frac{\Ve_{\rm
IG}(\sex)^{3/2}}{|\Ve_{\rm
IG,\se}(\sex)|}={1\over2\pi}\,\sqrt{\frac{\Vhi(\sgx)}{6\what\epsilon_\star}}
\simeq4.627\cdot 10^{-5},
\end{equation}
where the variables with subscript $\star$ are evaluated at
$\sg=\sgx$.

\paragraph{3.2.3}  The remaining inflationary observables (the spectral index $\ns$,
its running $\as$, and the tensor-to-scalar ratio $r$) --
estimated through the relations:
\beq\label{ns} \mbox{\ftn\sf (a)}\>\>\ns=\:
1-6\widehat\epsilon_\star\ +\ 2\widehat\eta_\star,~~~\mbox{\ftn\sf
(b)}\>\> \as =\:2\left(4\widehat\eta_\star^2-(n_{\rm
s}-1)^2\right)/3-2\widehat\xi_\star~~~ \mbox{and}~~~\mbox{\ftn\sf
(c)}\>\>r=16\widehat\epsilon_\star\, \eeq
with $\widehat\xi={\Ve_{\rm IG,\widehat\sg} \Ve_{\rm
IG,\widehat\sg\widehat\sg\widehat\sg}/\Ve_{\rm IG}^2}$ -- have to
be consistent with the data \cite{plin},  i.e.,
\begin{equation}  \label{nswmap}
\mbox{\ftn\sf
(a)}\>\>\ns=0.968\pm0.009~~~\mbox{and}~~~\mbox{\ftn\sf
(b)}\>\>r\leq0.12,
\end{equation}
at 95$\%$ \emph{confidence level} ({\sf\ftn c.l.}) -- pertaining
to the $\Lambda$CDM$+r$ framework with $|\as|\ll0.01$. Although
compatible with \sEref{nswmap}{b} the present combined \plk\ and
\bcp\ results \cite{gws} seem to favor $r$'s of order $0.01$ since
$r= 0.048^{+0.035}_{-0.032}$ at 68$\%$ c.l. has been reported.

\paragraph{3.2.4} Since SUGRA is an effective theory below $\mP=1$ the existence of higher-order terms in
$W$ and $K$, \eqs{Whi}{Kolg}, appears to be unavoidable.
Therefore, the stability of our inflationary solutions can be
assured if we entail
\beq \label{subP}\mbox{\ftn\sf (a)}\>\> \Vhi(\sgx)^{1/4}\leq1
~~~\mbox{and}~~~\mbox{\ftn\sf (b)}\>\>\sgx\leq1,\eeq
where the UV cutoff scale of the effective theory for the present
models is $\mP=1$, as shown in \Sref{fhi3}.

\begin{figure}[!t]\vspace*{-.44in}\begin{tabular}[!h]{cc}\begin{minipage}[t]{7.in}
\hspace*{.6in}
\epsfig{file=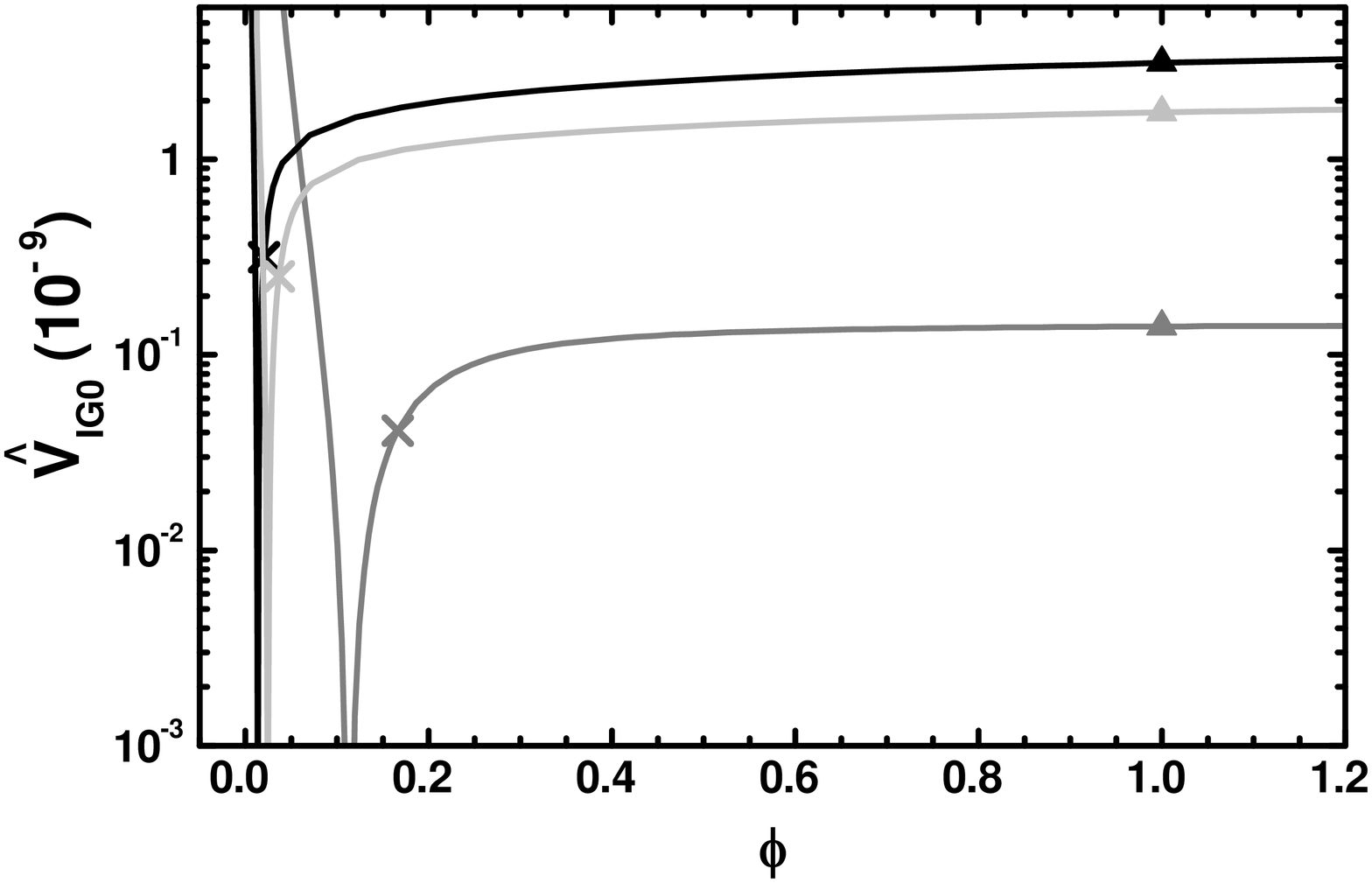,height=3.65in,angle=-90}\end{minipage}
&\begin{minipage}[h]{3.in}
\hspace{-3.5in}{\vspace*{-2.5in}\includegraphics[height=7.9cm,angle=-90]
{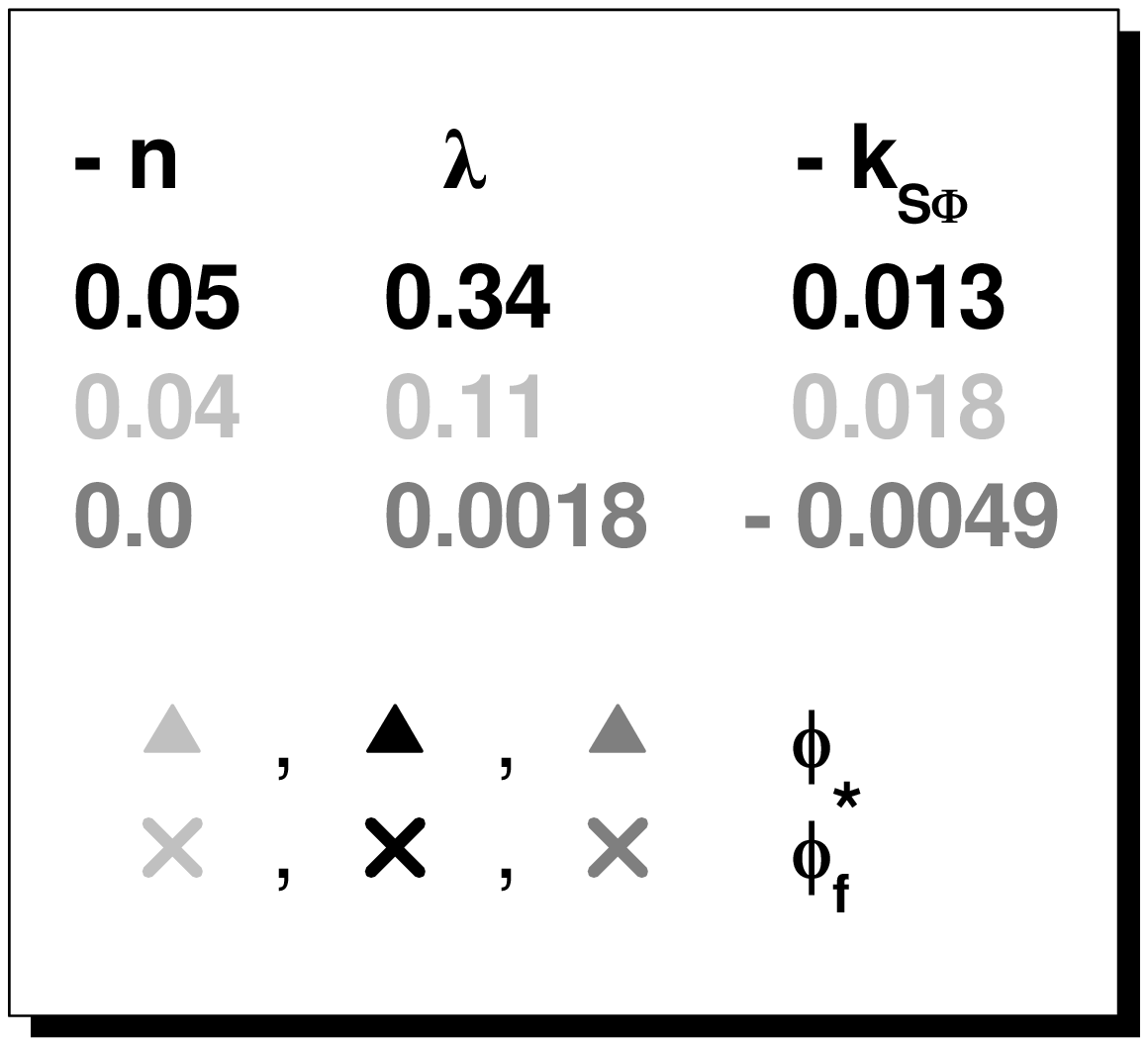}}\end{minipage}
\end{tabular}  \hfill \caption{\sl \small The inflationary
potential $\Vhio$  (gray, light gray and black line) as a function
of $\sg$ for $n=0,-1/25,-1/20$, $\ld=0.0013,0.11,0.34$ and
$\ksp\simeq0.0045,-0.018,-0.013$. Values corresponding to $\sgx$
and $\sgf$ are also depicted. }\label{fig3m}
\end{figure}

\paragraph{} The structure of $\Vhio$ as a function of $\sg$ for various $n$'s
is displayed in \Fref{fig3m}, where we depict $\Vhi$ versus $\sg$
imposing $\sgx=1$. The selected values of $\ld, \ksp$ and $n$,
shown in \Fref{fig3m}, yield $\ns=0.968$ and $r=0.0048, 0.061,
0.11$ for increasing $|n|$'s -- gray, light gray and  black line.
The corresponding $\ck$ values are $(0.078, 1.8,5.6)\cdot10^3$. We
remark that a gap of about one order of magnitude emerges between
$\Vhio(\sgx)$ -- and $\vev{\sg}$ -- for $|n|$ of order $0.01$ and
$n=0$ due to the larger $\ld$ and $\ck$ values employed for $n<0$;
actually, in the former case, $\Vhio^{1/4}(\sgx)$ -- and
$\vev{\phi}$ -- approaches the SUSY grand-unification scale,
$8.2\cdot10^{-3}$ -- cf. \cref{nmH}. This fact together with the
steeper slope that $\Vhio$ acquires close to $\sg=\sgx$ for $n<0$
is expected to have an imprint in elevating $\eph$ in \Eref{sr}
and, via \sEref{ns}{c}, on $r$.

\section{Results}\label{res}

Confronting our inflationary scenario with the requirements above
we can find its allowed parameter space.  We here present our
results for the two radically different cases: taking $n=0$ in
\Sref{res1} and $n<0$ in \Sref{res2}.

\subsection{$n=0$ Case}\label{res1}

We focus first on the form of \Ka\ induced by \Eref{Kolg} with
$n=0$. Our analysis in \Sref{res11} presents some approximate
expressions which assist us to interpret the numerical results
exhibited in \Sref{fgi2}.

\subsubsection{Analytic Results}\label{res11}

Upon substitution of \eqs{3Vhiom}{Jg} into \Eref{sr}, we can
extract the slow-roll parameters which determine the strength of
the inflationary stage. Performing expansions about $\xsg\simeq0$,
we can achieve approximate expressions which assist us to
interpret the numerical results presented below. Namely, we find
\beq \what\epsilon= \frac{(2 +\ksp \ck \xsg^4)^2}{3
\ft^2}~~~\mbox{and}~~~\what\eta={1\over 3 \ft^2}{\lf 4 + \ksp
\ck^2 \xsg^{6} + 2\ck\ksp\xsg^4-1\rg}.\eeq
%
As it may be numerically verified, the termination of \FHI is
triggered by the violation of the $\epsilon$ criterion at
$\sg=\sgf$, which does not decline a lot from its value for
$\ksp=0$. Namely we get
\beq \what\epsilon\lf\sgf\rg=1\>\Rightarrow\> \sgf=\sqrt{1 +
2/\sqrt{3}/\ck}. \label{sgap}\eeq
In the same approximation and given that $\sgf\ll\sgx$, $\Ns$ can
be calculated via \Eref{Nhi} with result
\beqs\beq \label{s*}
\Ns\simeq3\ck\lf{\sgx^2-\sgf^2}\rg/4\>\Rightarrow\>
\sgx\simeq2\sqrt{\Ns/3\ck}.\eeq
Obviously, \FHI with \sub\ $\sg$'s can be achieved if
\beq \label{fsub}
\sg_\star\leq1~~\Rightarrow~~\ck\geq4\Ns/3\simeq76 \eeq\eeqs
for $\Ne_\star\simeq52$. Therefore we need relatively large
$\ck$'s.

Replacing $\Vhio$ from \Eref{3Vhiom} in \Eref{Prob} we obtain
\begin{equation} \As^{1/2}=\frac{2\ld\ft^2(\sgx)}{8\sqrt{2}\pi\ck^2 \sgx^2(2 + \ksp
\ck \sgx^{4})}
\>\>\Rightarrow\>\>\ld\simeq2\pi\sqrt{2\As}\ck\lf{3\over\Ns}+{8\ksp\Ns\over3\ck}\rg\cdot
\label{lang} \eeq
Inserting finally \Eref{s*} into \sEref{ns}{a} and ({\sf\ftn c})
we can provide expressions for $n_{\rm s}$ and $r$. These are
\beq \label{gns}  \ns\simeq1-{2\over\what N_\star}\ +\
\frac{2\Ne_\star}{3\ck}\frac{32 \ksp +
27/\Ne_\star^3}{12}~~~\mbox{and}~~~ r\simeq {12\over\Ns^2}\ +\
64\, {4\ksp^2 \Ns^{2}\over 9\ck^{2}}\cdot\eeq
Therefore, a clear dependence of $\ns$ and $r$ on $\ksp$ arises,
with the first one being much more efficient. This depedence does
not exist within no-scale SUGRA since $\ksp$ vanishes by
definition -- see \Eref{nsks}.

\begin{figure}[!t]\vspace*{-.12in}
\hspace*{-.19in}
\begin{minipage}{8in}
\epsfig{file=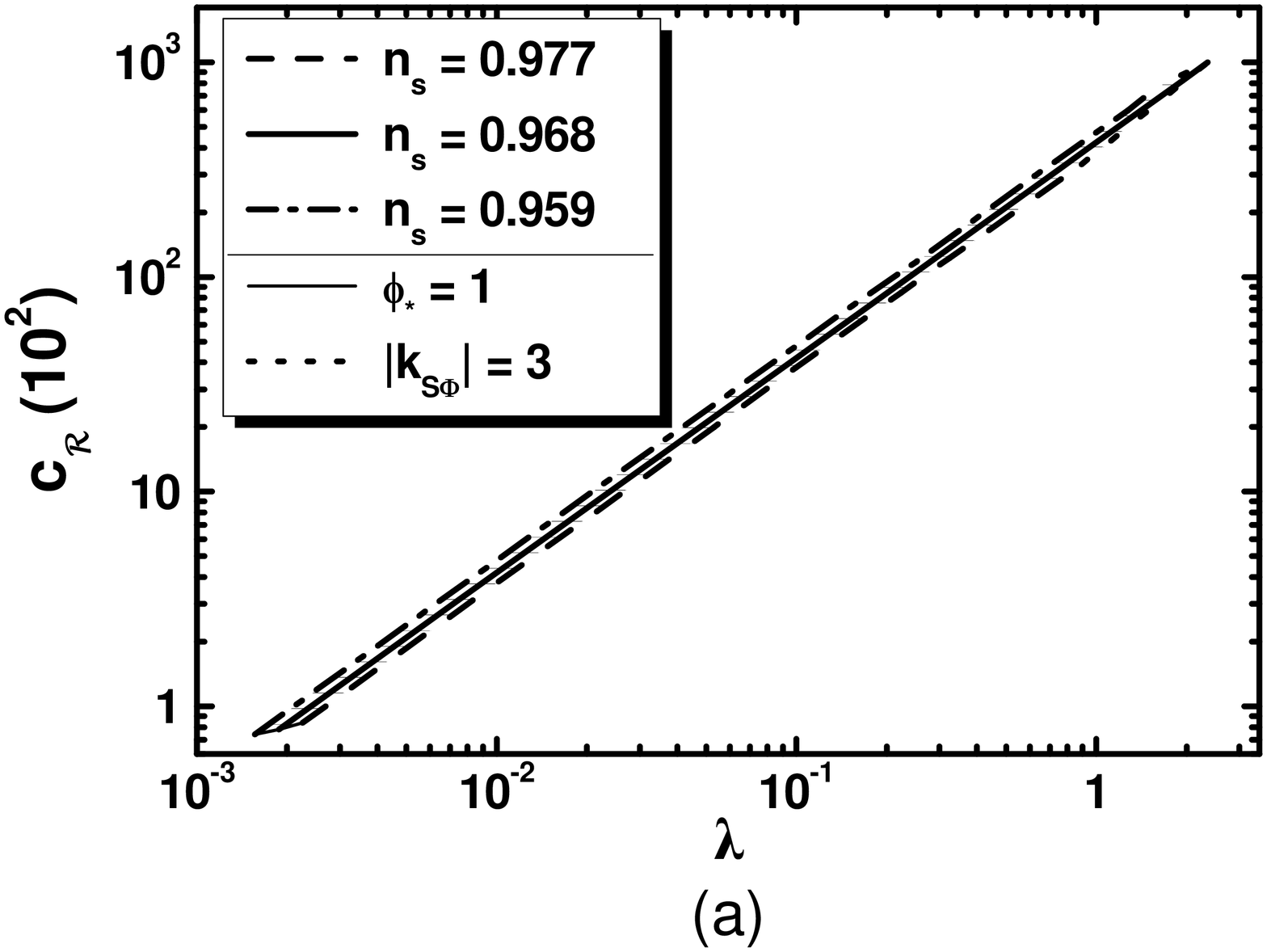,height=3.6in,angle=-90}
\hspace*{-1.2cm}
\epsfig{file=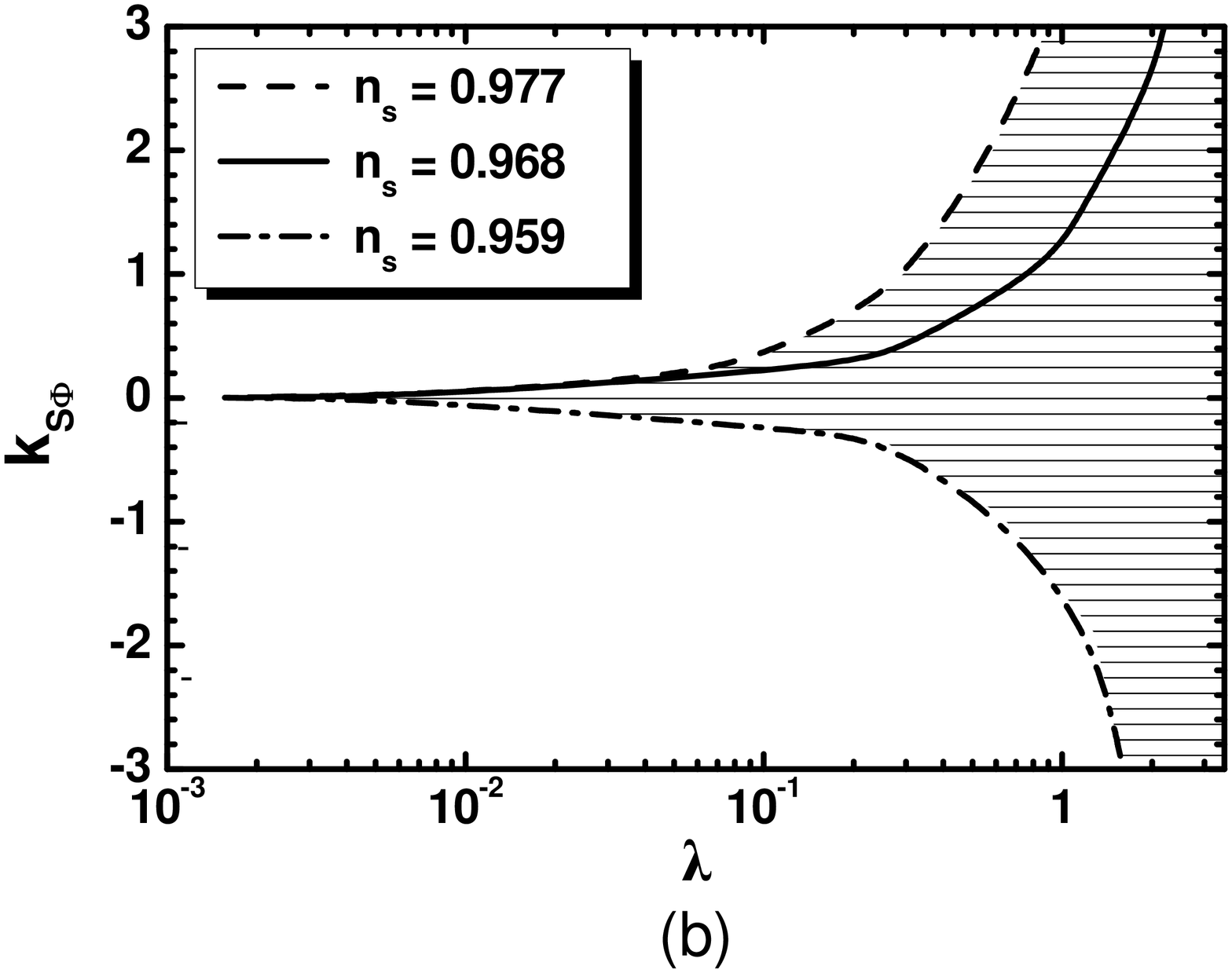,height=3.6in,angle=-90} \hfill
\end{minipage}
\hfill \caption{\sl\small Allowed (hatched) regions in the
$\ld-\ck$ plane (${\sf\ftn a}$) and $\ld-\ksp$ plane (${\sf\ftn
b}$) for $\ks=\kpp=0.5$. The conventions adopted for the various
lines are also shown.}\label{fig2g}
\end{figure}


\subsubsection{Numerical Results}\label{fgi2}

With fixed $\kns$ and $\Trh$ -- see Secs~\ref{fhim} and \ref{fhi2}
-- this inflationary scenario depends on the parameters:
\beq\ld,\>\ck,\>\kx,\>\ksp,\>\>\mbox{and}~~~\kpp.\label{para1}\eeq
Our results are independent of $\ks$, provided that $\what
m_{s}^2>0$ -- see in \Tref{tab4}. The same is also valid for
$\kpp\simeq1\ll\ck$ -- see \Eref{frsp}. We therefore set
$\kx=\kpp=0.5$. Besides these values, in our numerical code, we
use as input parameters $\ck,~\ksp$ and $\sgx$. For every chosen
$\ck\geq1$, we restrict $\ld$ and $\sgx$ so that the conditions
\eqss{Nhi}{Prob}{subP} are satisfied. By adjusting $\ksp$ we can
achieve $\ns$'s in the range of Eq.~(\ref{nswmap}). Our results
are displayed in \Fref{fig2g}-({\sf\ftn a}) and ({\sf\ftn b})
where we delineate the hatched regions allowed by the restrictions
of \Sref{fhi2} in the $\ld-\ck$ [$\ld-\ksp$] plane. The
conventions adopted for the various lines are also shown. In
particular, the dashed [dot-dashed] lines correspond to $n_{\rm
s}=0.977$ [$n_{\rm s}=0.959$], whereas the solid (thick) lines are
obtained by fixing $n_{\rm s}=0.968$ -- see Eq.~(\ref{nswmap}).
Along the thin line, which provides the lower bound for the
regions presented in \Fref{fig2g}, the constraint of
\sEref{subP}{b} is saturated. At the other end, the allowed
regions terminate along the dotted line where $|\ksp|=3$, since we
expect $\ksp$ values of order unity to be natural. From
\Fref{fig2g}-({\sf\ftn a}) we see that $\ck$ remains almost
proportional to $\ld$ and for constant $\ld$, $\ck$ increases as
$\ns$ decreases. From \Fref{fig2g}-({\sf\ftn b}) we remark that
$\ksp$ is confined close to zero for $n_{\rm s}=0.968$ and
$\ld<0.16$ or $\sgx>0.1$ -- see \Eref{s*}. Therefore, a degree of
tuning (of the order of $10^{-2}$) is needed in order to reproduce
the experimental data of \sEref{nswmap}{a}. On the other hand, for
$\ld>0.16$ (or $\sgx<0.1$), $\ksp$ takes quite natural (of order
one) negative values -- consistently with \Eref{gns}.

More explicitly, for $\ns=0.968$ and $\Ns\simeq52$ we find:
\bea\label{resg} && 78\lesssim
\ck\lesssim10^5~~~\mbox{with}~~~1.9\cdot10^{-3}\lesssim
\ld\lesssim2.35~~~\mbox{and}~~~ 0.005\lesssim \ksp\lesssim3\,.
\eea
Note that the former data dictated $\ksp<0$ since the central
$\ns$ was lower \cite{nIG}. Also we obtain $-7.8\lesssim
{\as/10^{-4}}\lesssim-7.4$ and $r\simeq4.4\cdot 10^{-3}$ which lie
within the allowed ranges of \Eref{nswmap}. On the other hand, the
results within no-scale SUGRA are much more robust since the
$\ksp$ (and $\kpp$) dependence collapses -- see \Eref{nsks}.
Indeed, no-scale SUGRA predicts identically
$\ns\simeq0.963,~\as=-6.5\cdot10^{-4}$ and $r=4\cdot10^{-3}$ which
are perfectly compatible with the data \cite{plin,gws} although
with low enough $r$.

\subsection{$n<0$ Case}\label{res2}

Following the strategy of the previous section, we present below
first some analytic results in \Sref{fgin1}, which provides a
taste of the numerical findings exhibited in \Sref{fgin2}.

\subsubsection{Analytic Results}\label{fgin1}

Plugging \eqs{3Vhiom}{Jg} into \Eref{sr} and taking $\kpp\simeq0$,
we obtain the following approximate expressions for the slow-roll
parameters
\bea \nonumber &\what\epsilon&=\frac{(2 + 3 n - 3n\ck \xsg^2 + (1
+ 3 n)\ksp \ck\xsg^4)^2}{3 (1 + n) \fsp^2
\fpp^2}~~~\mbox{and}~~~\what\eta=\frac{1}{3 (1 + n) \fsp^2
\fpp^2}\times\\&\times& 2 \Big[\xsg^2\Big(\ksp \lf\xsg^2 \lf 6 \ck
+ \ck^2\xsg^2 + \ksp \ck^2 \xsg^4\rg-11\rg-2 \ck \Big)+ 9 n^2
\fsp^2 \fpp^2\nonumber\\&+ & 4  + 6 n \fsp \fpp \lf2 + \ksp \xsg^2
(\ck \xsg^2-3)\rg\Big] \,.\label{gmsr1}\eea
Taking the limit of the expressions above for $\ksp\simeq0$ we can
analytically solve the condition in \Eref{sr} w.r.t $\xsg$. The
results are
\beq {\sg_{1\rm
f}}=\sqrt{\frac{3(1-n)+2\sqrt{3(1+n)}}{3(1+n)\ck}}~~~\mbox{and}~~~{\sg_{2\rm
f}}=\sqrt{\frac{1-9n+\sqrt{16+21n(3n-1)}}{3(1+n)\ck}}\,\cdot\label{sgf}\eeq
The end of IGI mostly occurs at $\sgf=\sg_{1\rm f}$ because this
is mainly the maximal value of the two solutions above. Since
$\sgf\ll\sgx$, we can estimate $\Ns$ through \Eref{Nhi} neglecting
$\sgf$. Our result is
\beqs\beq \Ns \simeq  (1 + n) \frac{3 n \ln\sgx + \ln\lf 2 + 3 n -
3 \ck n \sgx^2\rg}{|n|(2 + 3 n)}\cdot\label{Ngm}\eeq
Ignoring the first term in the last equality and solving w.r.t
$\xst$ we extract $\sgx$ as follows -- cf. \cref{nIG,nMCI}:
\beq\label{sm*}\sgx\simeq\sqrt{(2-\re)/3n \ck}~~~\mbox{with}~~~\re
= e^{-n (2 + 3 n)\Ns/(1 + n)}\,.\eeq
Although a radically different dependence of $\sgx$ on $\Ns$
arises compared to the model of \Sref{res1} -- cf. \Eref{s*} --
$\sgx$ can again remain \sub\ for large $\ck$'s. Indeed,
\beq \label{fmsub} \sgx\leq1~~~\Rightarrow~~~\ck\geq
(2-\re)/3n\,.\eeq\eeqs
On the other hand, $\se_\star$ remains \trns, since plugging
\Eref{sm*} into \Eref{se1} we find
\beq \se_\star\simeq-\sqrt{3(1 + n)/2}\lf {4(2+3n)\Ns/(1 + n)}+\ln
3|n|\rg, \label{sem*}\eeq
which gives $\se_\star=7-10$ for $\se_{\rm c}=0$ and
$n=-(0.03-0.05)$ -- independently of $\ck$. Despite this fact, our
construction remains stable under possible corrections from
non-renormalizable terms in $\fk$ since these are expressed in
terms of initial field $\Phi$, and can be harmless for
$|\Phi|\leq1$.

Upon substitution of \Eref{sm*} into \Eref{Prob} we end up with
\begin{equation} \ld \simeq\frac{4\pi \sqrt{2\ck\As}\re^{3n/2}(\re-2)
(\ksp (1 + 3 n) (\re-2)^2+9n^2\ck \re)}{3 |n|^{(3n+1)/2} \sqrt{(1
+ n)(\ksp(\re-2)-3n\ck}}\cdot \label{langm} \eeq
We remark that $\ld$ depends not only on $\ck$ and $\ksp$ as in
\Eref{lang} but also on $n$. Inserting \Eref{sm*} into
\Eref{gmsr1}, employing then \sEref{ns}{a} and expanding for
$\ck\gg1$ we find
\beqs\beq \label{nsgm} \ns=\frac{(\re-2)^2+n(\re-2)(\re+12)
-6n^2\re^2}{(1 + n) (\re+3n-2)^2}\ +\
4\ksp\,\re\,\frac{\re(4-(1+3m)\re)-4}{9n(1 + n)(\re+3
n-2)^2\ck}\cdot \eeq
Following the same steps, from \sEref{ns}{c} we find
\beq \label{ragm} r=16\lf\frac{3n^2\re^2}{(1 + n) (\re+3 n-2)^2}\
+\ 2\ksp\re  \frac{4+(\re-3n\re - 4)\re}{3(1 + n)(\re+3 n-
2)^2\ck}\rg\cdot\eeq\eeqs
From the above expressions we see that primarily $|n|\neq0$ and
secondarily $n<0$ help to sizably increase $r$. Given that
$\re\gg1$, $\ns$ is close to unity as can be infered by the first
ratio in the right-hand side of \Eref{nsgm}. Any increase of $\ns$
due to the existence $|n|\neq0$ can be balanced by a choise of
$\ksp<0$. Note that the second term in \Eref{nsgm} is less
suppressed w.r.t the second term in \Eref{ragm} since $\ck\gg1$ is
multiplied by $n\ll1$.

\subsubsection{Numerical Results} \label{fgin2}

Besides the free parameters shown in \Eref{para1} we have also
here $n$, which is constrained to negative values. Using the
reasoning explained in \Sref{fgi2} we set $\kpp=0.5$. On the other
hand, $\what m_{s}^2$ can become positive with $\ks$ lower than
the value used in \Sref{fgi2} since positive contributions from
$n<0$ arises here -- see \Tref{tab4}. Moreover, if $\ks$ takes a
value of order unity $\what m_{s}^2$ grows more efficiently than
in the case with $n=0$, rendering thereby the RCs in \Eref{Vhic}
sizeable for very large $\ck$ values ($\sim10^5$). To avoid such
dependence of the model predictions on the RCs, we use $\kx$
values lower than those used in \Sref{fgi2}. Thus, we set
$\kx=0.05$ throughout. As in the previous case,
\eqss{Nhi}{Prob}{subP} assist us to restrict $\ld$ (or $\ck\geq1$)
and $\sgx$. By adjusting $n$ and $\ksp$ we can achieve not only
$\ns, \as$ and $r$ values in the range of \Eref{nswmap} but also
$r$'s close to the central value reported in \cref{gws}.

Confronting the parameters with Eqs.~(\ref{Nhi}), (\ref{Prob}),
(\ref{nswmap}{\sffamily\ftn a, b}) and (\ref{subP}) we depict the
allowed (hatched) regions in the $\ld-\ck$, $\ld-\ksp$, $\ld-r$
and $\ld-\as$ planes for $n=-1/30$ (gray lines and hatched
regions), $n=-1/25$ (light gray lines and hatched regions),
$n=-1/20$ (black lines and hatched regions) in \sFref{fig2gm}{a},
{\sf\ftn (b), (c)} and {\sf\ftn (d)} respectively. Note that the
conventions adopted for the various lines are identical with those
used in \Fref{fig2g} -- i.e., the dashed, solid (thick) and
dot-dashed lines correspond to $\ns=0.977, 0.968$ and $0.959$
respectively, whereas along the thin (solid) lines the constraint
of \sEref{subP}{b} is saturated. The perturbative bound on $\ld$
limits the various regions at the other end.

\begin{figure}[!t]\vspace*{-.12in}
\hspace*{-.19in}
\begin{minipage}{8in}
\epsfig{file=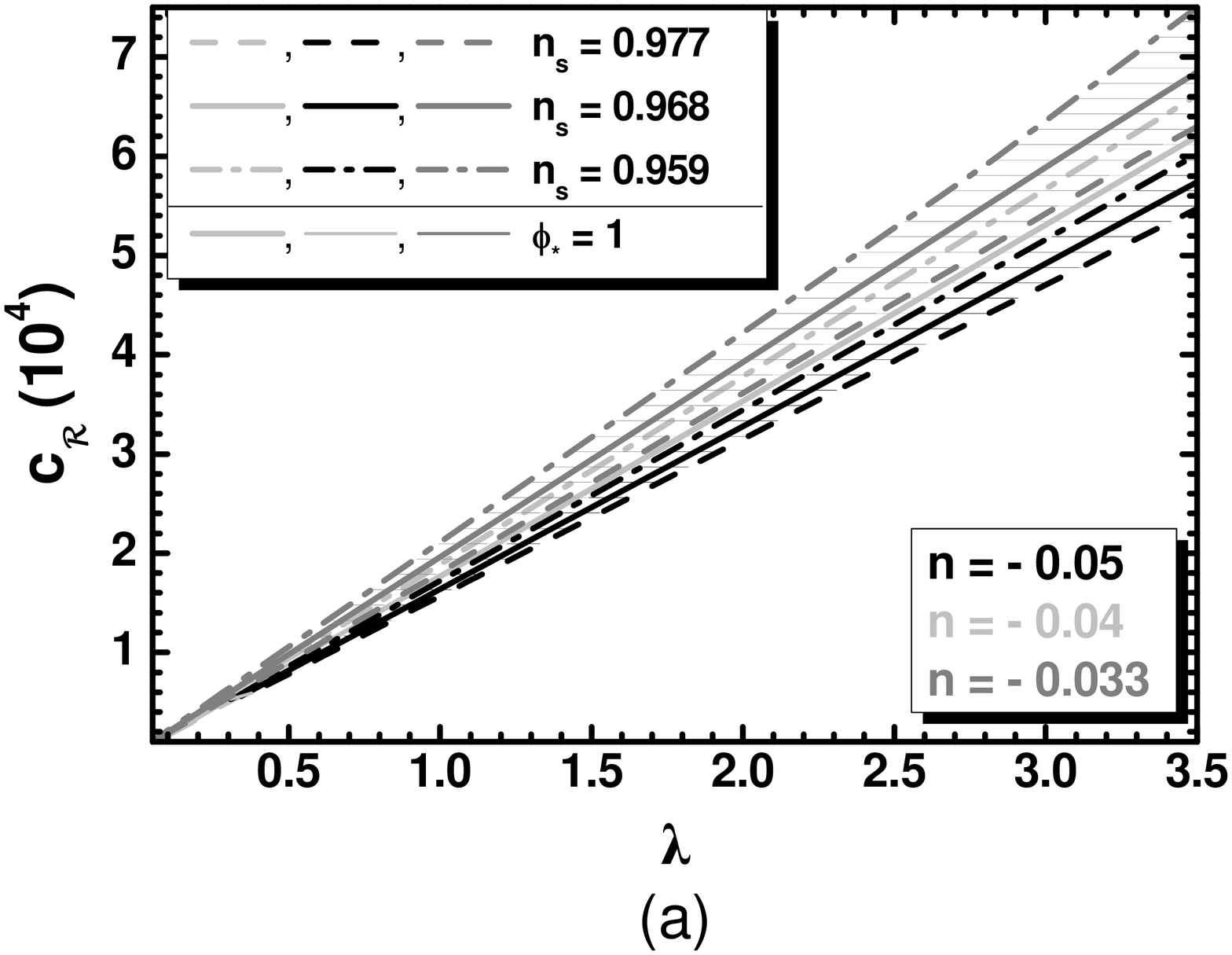,height=3.6in,angle=-90}
\hspace*{-1.2cm}
\epsfig{file=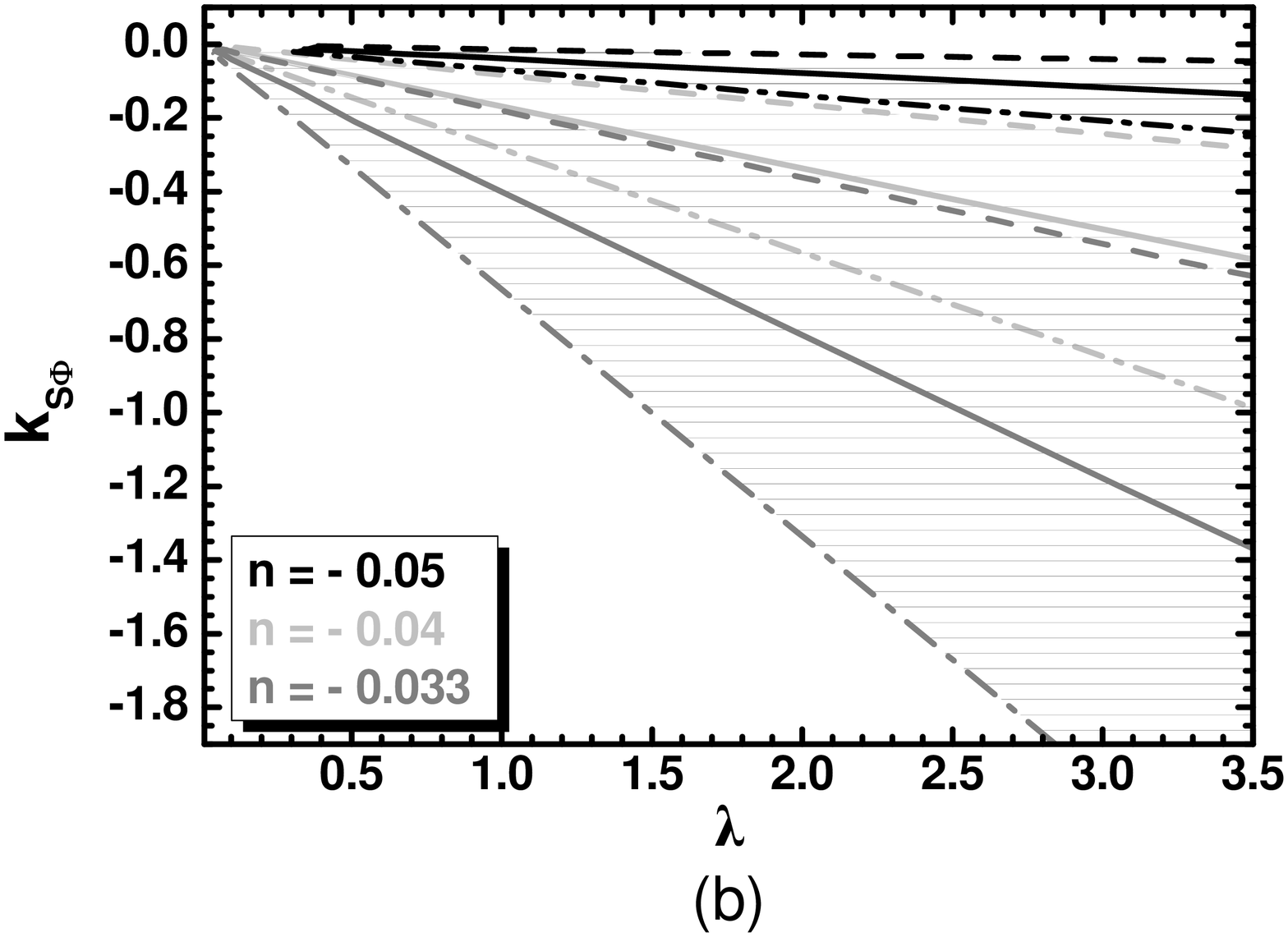,height=3.6in,angle=-90} \hfill
\end{minipage}
\hfill \hspace*{-.19in}
\begin{minipage}{8in}
\epsfig{file=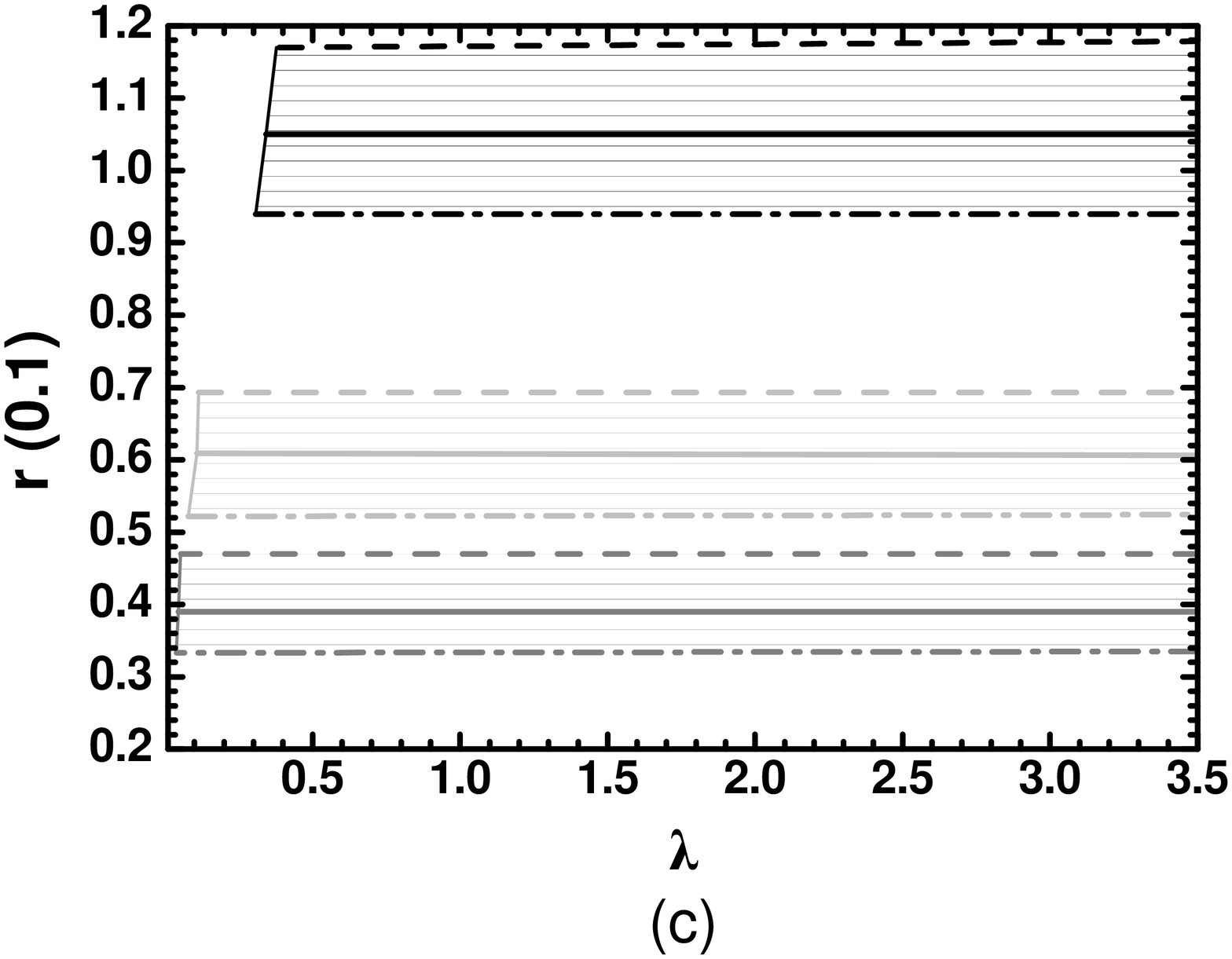,height=3.6in,angle=-90}
\hspace*{-1.2cm}
\epsfig{file=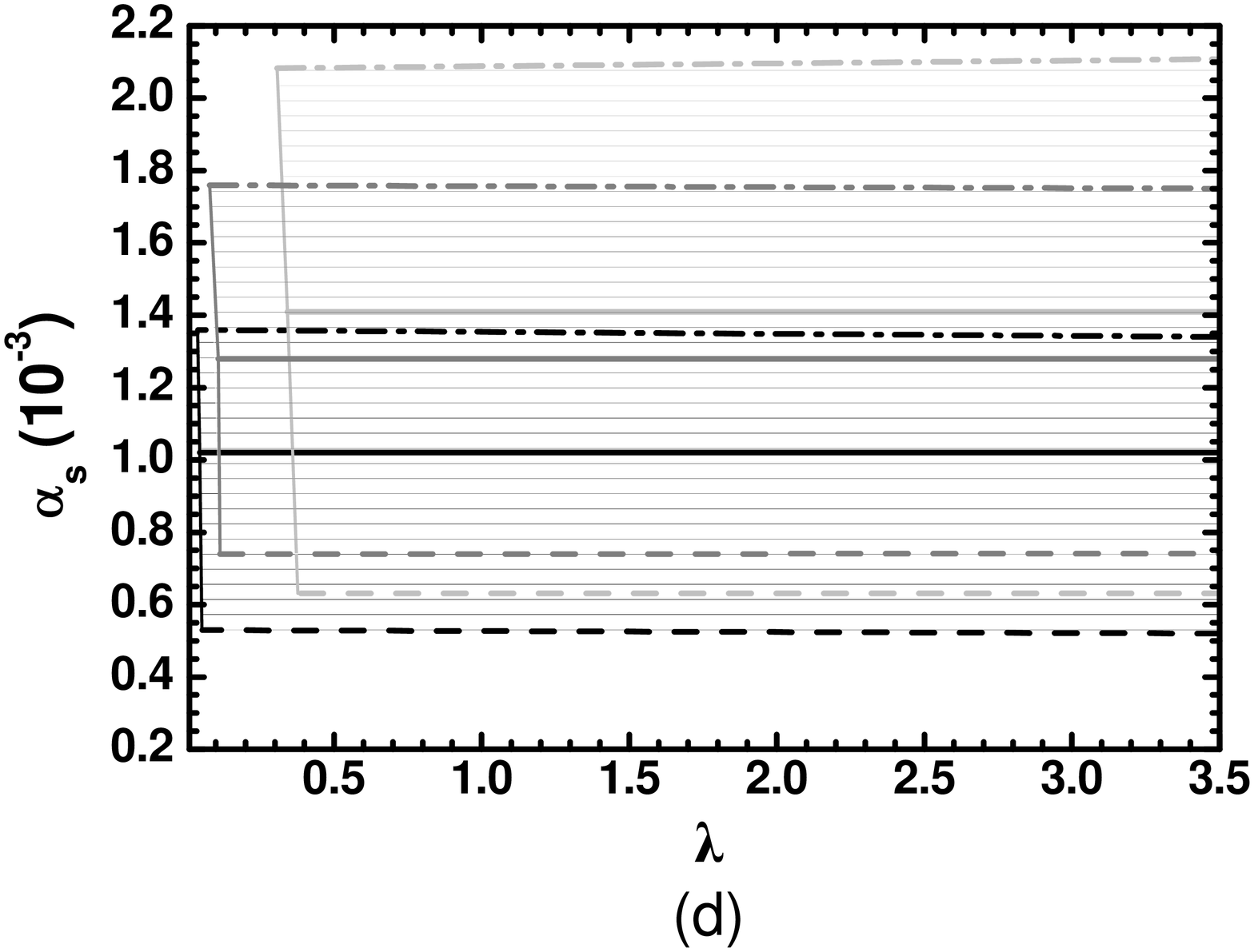,height=3.6in,angle=-90} \hfill
\end{minipage}
\hfill \caption{\sl\small Allowed (hatched) regions in the
$\ld-\ck$ ({\sf\ftn a}), $\ld-\ksp$ ({\sf\ftn b}), $\ld-r$
({\sf\ftn c}), $\ld-\as$ ({\sf\ftn d}) plane for $\ks=0.1$,
$\kpp=0.5$ and $n=-0.033$ (gray lines and hatched regions),
$n=-0.04$ (light gray lines and hatched regions), $n=-0.05$ (black
lines and hatched regions). The conventions adopted for the type
and color of the various lines are also shown in the label of
panel {\sf\ftn (a)}.}\label{fig2gm}
\end{figure}


From \sFref{fig2gm}{a} we remark that $\ck$ remains almost
proportional to $\ld$ but the dependence on $\ksp$ is stronger
than that shown in \sFref{fig2g}{a}. Also, as $|n|$ increases, the
allowed areas are displaced to larger $\ld$ and $\ck$ values in
agreement with \Eref{fmsub} -- cf. \Fref{fig2g}. Similarly, the
allowed $\ksp$'s move to larger values as $|n|$ and/or $\ns$
increases. For fixed $\ns$, increasing $\ck$ entails a decrease of
$\ksp$ in accordance with \Eref{nsgm}. Finally, from
\sFref{fig2gm}{c} and {\sf\ftn (d)} we conclude that employing
$|n|\gtrsim0.01$, $r$ and $\as$ increase w.r.t their values for
$n=0$ -- see results below \Eref{resg}. As a consequence,  for
$n\simeq-(0.03-0.05)$, $r$ enters the observable region. On the
other hand, $\as$ although one order larger than its value for
$n=0$ remains sufficiently low; it is thus consistent with the
fitting of data with the standard $\Lambda$CDM$+r$ model -- see
\Eref{nswmap}. As anticipated below \Eref{ragm}, the resulting
$r$'s depend only on the input $n$ and $\ksp$ (or $\ns$), and are
independent of $\ld$ (or $\ck$). The same behavior is also true
for $\as$. It is worth noticing that the existence of $\ksp\neq0$
is imperative for the viability of our scheme.  More explicitly,
for $n_{\rm s}=0.968$ and $\Ne_\star\simeq55-57$ we find:
\beqs\bea\label{resgm} && 0.09\lesssim {\ck\over
10^4}\lesssim6.9~~~\mbox{with}~~~0.045\lesssim
\ld\lesssim3.5~~~\mbox{and}~~~ 0.18\lesssim
-{\ksp\over0.1}\lesssim10.4\, ~~~(n=-0.033);~~~~~~\\ &&
0.19\lesssim {\ck\over
10^4}\lesssim6.7~~~\mbox{with}~~~0.11\lesssim
\ld\lesssim3.5~~~\mbox{and}~~~ 0.18\lesssim
-{\ksp\over0.1}\lesssim6.3\, ~~~(n=-0.04);~~~~~~\\ &&
\label{resgm2} 0.56\lesssim {\ck\over
10^4}\lesssim6.1~~~\mbox{with}~~~0.34\lesssim
\ld\lesssim3.5~~~\mbox{and}~~~ 0.13\lesssim
-{\ksp\over0.1}\lesssim1.45\,~~~(n=-0.05).~~~~~~~~~~~~\> \eea\eeqs
In these regions we obtain
\beq \label{resgm4}  {r\over0.1}=0.4, 0.6, 1.05~~~\mbox{and}~~~
{\as\over0.001}=1,1.3,1.4~~~\mbox{for}~~~
-{n\over0.01}=3.3,4,5\eeq respectively. It is impressive that the
observable $r$'s above are achieved with \sub\ $\phi$'s. However,
this fact does not contradict to the Lyth bound \cite{lyth}, since
this is applied to the (totally auxiliary) EF inflaton $\se$ which
remains \trns -- see \Eref{sem*}.

\section{Effective Cut-off Scale}\label{fhi3}

An outstanding trademark of IGI is that it is unitarity-safe
\cite{nIG, R2r, gian}, despite the fact that its implementation
with \sub\ $\phi$'s -- see \eqs{fsub}{fmsub} -- requires
relatively large $\ck$'s. To show this, we below extract the UV
cut-off scale, $\Qef$, of the effective theory first in the JF --
see \Sref{fhi3b} -- and then in the EF -- see \Sref{fhi3a}.

\subsection{Jordan Frame Computation}\label{fhi3b}

If we expand $g_{\mu\nu}$ about the flat spacetime metric
$\eta_{\mu\nu}$ and $\phi$ about its v.e.v as follows
\beq\label{vevs1}
g_{\mu\nu}\simeq\eta_{\mu\nu}+{h_{\mu\nu}}~~~\mbox{and}~~~\phi=\vev{\phi}
+\dph~~~\mbox{with}~~~\vev{\phi}=1/\sqrt{\ck}\eeq
-- where $h^{\mu\nu}$ is the graviton --, the lagrangian
corresponding to the two first terms in the right-hand side of
${\sf S}$ in \Eref{Sfinal} for $\al=\Phi$ takes the form -- cf.
\cref{cutof}:
\bea \nonumber \delta{\cal L}&=&-{\vev{\fr}\over4}{F}_{\rm EH}\lf
h^{\mu\nu}\rg +\frac{\vev{F_{\rm K}}}{2}\partial_\mu
\dph\partial^\mu\dph+\lf \vev{f_{
R,\phi}}\dph+\frac12\vev{f_{R,\phi\phi}}\dph^2+\cdots\rg F_{R}\\
&=&-{1\over8}F_{\rm EH}\lf \bar h^{\mu\nu}\rg+ \frac12\partial_\mu
\overline\dph\partial^\mu\overline\dph+\Qef^{-1}\overline\dph^{2}\,\Box
\bar h\,,\label{jf1}\eea
where the functions $F_{\rm EH}$ and $F_{R}$ related to the the
linearized Einstein-Hilbert part of the lagrangian, read
\beq {F}_{\rm EH}\lf h^{\mu\nu}\rg= h^{\mu\nu} \Box
h_{\mu\nu}-h\Box h+2\partial_\rho h^{\mu\rho}\partial^\nu
h_{\mu\nu}-2\partial_\nu h^{\mu\nu}\partial_\mu
h~~~\mbox{and}~~~F_{R}\lf h^{\mu\nu}\rg=\Box h-
\partial_\mu \partial_\nu h^{\mu\nu}\,\label{Fr}\eeq
with $h=h^\mu_{\mu}$. Also $F_{\rm K}$ along the trajectory in
\Eref{inftr} is calculated to be
\beq \label{Fk} F_{\rm
K}=\Omega_{\Phi{\Phi^*}}-\frac{n\Omega_{\Phi}\Omega_{\Phi^*}}{(1+n)\Omega}=\frac{\kns}{1+n}+6n\ck\,\cdot\eeq
Moreover, $\bar h_{\mu\nu}$ and $\overline\dph$ are the JF
canonically normalized fields defined by the relations
\beq \overline\dph=\sqrt{\frac{\vev{\fr}}{\vev{\bar f_{
R}}}}\dph~~~\mbox{and}~~~ \bar h_{\mu\nu}=
\sqrt{\vev{\fr}}\,h_{\mu\nu}+\frac{\vev{f_{
R,\phi}}}{\sqrt{\vev{\fr}}}\eta_{\mu\nu}\dph ~~~\mbox{with}~~~\bar
f_{R}=F_{\rm K}\fr+\frac32 f_{R,\phi}^2\,.\label{Jcan}\eeq
Finally, $\Qef$ in \Eref{jf1} is the JF UV cut-off scale since it
controls the strength of the $\dph-\dph$ scattering process via
$s$-channel $h^{\mu\nu}$ exchange. It is determined via the
relation
\beq \label{Luv} \Qef=  \frac{2\vev{\bar
f_{R}}}{\sqrt{\vev{\fr}}\vev{f_{R,\phi\phi}}}\simeq
\frac{6(1+n)}{\sqrt{1-{\kns\over6(1+n)\ck}}}\,\cdot\eeq
For the estimations above we make use of \eqs{vevs1}{Fk}. Since
the dangerous factor $\ck^{-1}$ included in $\vev{f_{R,\phi\phi}}$
is eliminated in \Eref{Luv}, the theory can be characterized as
unitarity-safe.

\subsection{Einstein Frame Computation}\label{fhi3a}

Alternatively, $\Qef$ can be determined in EF, following the
systematic approach of \cref{riotto}. At the SUSY vacuum in
\Eref{vevs}, the EF (canonically normalized) inflaton is found via
\Eref{se1} to be
\beq\dphi=\vev{J}\dph~~~\mbox{with}~~~\vev{J}\simeq\sqrt{6(1+n)\ck}\,.
\label{dphi} \eeq
The fact that $\dphi$ does not coincide with $\dph$ at the vacuum
of the theory -- contrary to the standard Higgs non-minimal
inflation \cite{cutoff} -- ensures that our models are valid up to
$\mP=1$. To show it, we write ${\sf S}$ in \Eref{Saction1} along
the path of \Eref{inftr} as follows
 \beq\label{S3} {\sf S}=\int d^4x \sqrt{-\what{
\mathfrak{g}}}\lf-\frac{1}{2} \rce +\frac12\,J^2
\dot\phi^2-\Ve_{\rm IG0}+\cdots\rg, \eeq
where the ellipsis represents terms irrelevant for our analysis;
$J$ and $\Vhio$ are given by \eqs{Jg}{3Vhiom} respectively. We
first expand $J^2 \dot\phi^2$ about $\vev{\phi}$ in terms of
$\dphi$ in \Eref{dphi} and we arrive at the following result
\beqs\beq\label{exp2} J^2
\dot\phi^2=\lf1-\sqrt{\frac{2}{3(1+n)}}{\dphi}+\frac{\dphi^2}{2(1+n)}-
\sqrt{\frac{2}{3(1+n)^3}}\frac{\dphi^3}{3}+\cdots\rg\dot\dphi^2.\eeq
The expansion corresponding to $\Vhio$ in \Eref{3Vhiom}  with
$\ksp\simeq0$ and $\kpp\simeq0$ includes the terms:
\beq
\Vhio=\frac{\ld^2\dphi^2}{6\ck^4(1+n)}\lf1+\frac{\dphi}{\sqrt{6(1+n)^3}}+
\frac{\dphi^2}{24(1+n)^2}-\cdots\rg\cdot~~~\label{Vexp}\eeq\eeqs
From Eqs.~(\ref{exp2}) and (\ref{Vexp}) we conclude that $\Qef=1$,
in agreement with our analysis in \Sref{fhi3b}.

\section{Conclusions}\label{con}

We updated the analysis of IGI introduced in \cref{nIG}, in the
view of the combined recent analysis of the \plk\ and \bcp\
results \cite{plin, gws}. These inflationary models are tied to a
superpotential, which realizes easily the idea of induced gravity,
and a logarithmic \Ka, which includes all the allowed terms up to
the fourth order in powers of the various fields -- see
\Eref{Kolg}. We also allowed for deviations from the prefactor
$(-3)$ multiplying the logarithm of the \Ka, parameterizing it by
a factor $(1+n)$. The models are totally defined imposing two
global symmetries -- a continuous $R$ and a discrete
$\mathbb{Z}_2$ symmetry -- in conjunction with the requirement
that the original inflaton takes \sub\ values.

In the case of no-scale SUGRA, thanks to the underlying
symmetries, the inflaton is not mixed with the accompanying
non-inflaton field in the \Ka. As a consequence, the model
predicts $\ns\simeq0.963$, $\as\simeq-0.00065$ and $r\simeq0.004$,
in excellent agreement with the current \plk\ data. Beyond
no-scale SUGRA, for $n=0$, we showed that $\ns$ spans the entire
allowed range in \sEref{nswmap}{a} by conveniently adjusting the
coefficient $\ksp$. In addition, for $n\simeq-(0.03-0.05)$, $r$
becomes compatible with the 1-$\sigma$ domain of the joint
analysis of \plk\ and \bcp\ data and accessible to the ongoing
measurements with negligibly small $\as$. In this last case a mild
tuning of $\kx$ to values of order $0.05$ is adequate so that the
one-loop RCs remain subdominant. Moreover, in all cases, the
corresponding effective theory is valid up to the Planck scale.

\acknowledgments This research was supported from the MEC and
FEDER (EC) grants FPA2011-23596 and the Generalitat Valenciana
under grant PROMETEOII/2013/017.

\def\ijmp#1#2#3{{\emph{Int. Jour. Mod. Phys.}}
{\bf #1},~#3~(#2)}
\def\plb#1#2#3{{\emph{Phys. Lett.  B }}{\bf #1},~#3~(#2)}
\def\zpc#1#2#3{{Z. Phys. C }{\bf #1},~#3~(#2)}
\def\prl#1#2#3{{\emph{Phys. Rev. Lett.} }
{\bf #1},~#3~(#2)}
\def\rmp#1#2#3{{Rev. Mod. Phys.}
{\bf #1},~#3~(#2)}
\def\prep#1#2#3{\emph{Phys. Rep. }{\bf #1},~#3~(#2)}
\def\prd#1#2#3{{\emph{Phys. Rev.  D} }{\bf #1},~#3~(#2)}
\def\npb#1#2#3{{\emph{Nucl. Phys.} }{\bf B#1},~#3~(#2)}
\def\npps#1#2#3{{Nucl. Phys. B (Proc. Sup.)}
{\bf #1},~#3~(#2)}
\def\mpl#1#2#3{{Mod. Phys. Lett.}
{\bf #1},~#3~(#2)}
\def\arnps#1#2#3{{Annu. Rev. Nucl. Part. Sci.}
{\bf #1},~#3~(#2)}
\def\sjnp#1#2#3{{Sov. J. Nucl. Phys.}
{\bf #1},~#3~(#2)}
\def\jetp#1#2#3{{JETP Lett. }{\bf #1},~#3~(#2)}
\def\app#1#2#3{{Acta Phys. Polon.}
{\bf #1},~#3~(#2)}
\def\rnc#1#2#3{{Riv. Nuovo Cim.}
{\bf #1},~#3~(#2)}
\def\ap#1#2#3{{Ann. Phys. }{\bf #1},~#3~(#2)}
\def\ptp#1#2#3{{Prog. Theor. Phys.}
{\bf #1},~#3~(#2)}
\def\apjl#1#2#3{{Astrophys. J. Lett.}
{\bf #1},~#3~(#2)}
\def\n#1#2#3{{Nature }{\bf #1},~#3~(#2)}
\def\apj#1#2#3{{Astrophys. J.}
{\bf #1},~#3~(#2)}
\def\anj#1#2#3{{Astron. J. }{\bf #1},~#3~(#2)}
\def\mnras#1#2#3{{MNRAS }{\bf #1},~#3~(#2)}
\def\grg#1#2#3{{Gen. Rel. Grav.}
{\bf #1},~#3~(#2)}
\def\s#1#2#3{{Science }{\bf #1},~#3~(#2)}
\def\baas#1#2#3{{Bull. Am. Astron. Soc.}
{\bf #1},~#3~(#2)}
\def\ibid#1#2#3{{\it ibid. }{\bf #1},~#3~(#2)}
\def\cpc#1#2#3{{Comput. Phys. Commun.}
{\bf #1},~#3~(#2)}
\def\astp#1#2#3{{Astropart. Phys.}
{\bf #1},~#3~(#2)}
\def\epjc#1#2#3{{Eur. Phys. J. C}
{\bf #1},~#3~(#2)}
\def\nima#1#2#3{{Nucl. Instrum. Meth. A}
{\bf #1},~#3~(#2)}
\def\jhep#1#2#3{{\emph{JHEP} }
{\bf #1},~#3~(#2)}
\def\jcap#1#2#3{{\emph{JCAP} }
{\bf #1},~#3~(#2)}

\newcommand{\hepth}[1]{{\tt hep-th/#1}}
\newcommand{\hepph}[1]{{\tt hep-ph/#1}}
\newcommand{\hepex}[1]{{\tt hep-ex/#1}}
\newcommand{\astroph}[1]{{\tt astro-ph/#1}}
\newcommand{\arxiv}[1]{{\tt arXiv:#1}}

\end{document}